\documentclass[aps, prd, reprint, superscriptaddress, nofootinbib, longbibliography, showkeys]{revtex4-2}
\IfFormatAtLeastTF{2023-06-01}{\let\label\LatexKernelLabel}{}

\usepackage{amsmath}
\usepackage{amssymb}
\usepackage{bm}
\usepackage{mathtools}
\usepackage{graphicx}
\usepackage{dcolumn}
\usepackage{booktabs}
\usepackage[stretch=35]{microtype}
\usepackage{hyperref}

\newcommand{\Order}{\mathcal{O}}

\begin{document}

\title{Scalar decoupling and critical-curve invariance in a reduced MOG black-hole theory}

\author{Nikko John Leo S. Lobos}
\email{nikko\_john\_s\_lobos@dlsu.edu.ph}
\affiliation{Department of Physics, De La Salle University, 2401 Taft Ave, Malate, Manila, 1004 Metro Manila, Philippines}

\author{Emmanuel T. Rodulfo}
\email{emmanuel.rodulfo@dlsu.edu.ph}
\affiliation{Department of Physics, De La Salle University, 2401 Taft Ave, Malate, Manila, 1004 Metro Manila, Philippines}

\date{\today}

\begin{abstract}
Linear perturbations of a Schwarzschild-MOG black hole separate into scalar and gravito-vector sectors in the reduced strong-field theory at leading order in $\mu\bar G M\ll1$. The field redefinition $\widehat h_{\mu\nu}=h_{\mu\nu}-\psi\bar g_{\mu\nu}$ gives a decoupled Klein--Gordon equation for the scalar and the Reissner--Nordstr\"om equations for the gravito-vector sector. The scalar is stable under the stated spectral boundary conditions for $m_G^2\ge0$. A scalar-only excitation produces a conformal metric perturbation with null Hamiltonian contribution $H_1^{(s)}=-\psi H_0$. For fixed observer worldlines and conformally corresponding orthonormal frames, every linear scalar-to-critical-curve transfer coefficient vanishes, including monopole breathing, provided the conformal factor is smooth, positive, and preserves the capture boundary. The exceptional charged axial dipole instead induces a constrained metric perturbation. A first-order radial integral reconstructs this perturbation in a gauge regular at the future horizon and on the sphere, separating the radiative amplitude from stationary angular momentum without division by frequency. For a unit-incident canonical vector amplitude with zero angular momentum, matched asymptotic solutions give an explicit charge-dependent local metric response that vanishes as $\omega^2$ at fixed radius and fixed subextremal background. A separate infinitesimal Kerr--Newman benchmark yields the stationary impact-parameter midpoint shift and recovers $\delta C_b=-2a_{\rm rot}$ in the neutral limit. These results exclude linear scalar breathing of the specified angular critical curve and distinguish the low-frequency vector-induced metric response from a stationary rotation effect.
\end{abstract}


\maketitle

\section{Introduction}
\label{sec:introduction}

Black-hole ringdown and horizon-scale lensing probe the same strong-field spacetime through different physical processes. Ringdown depends on the propagating degrees of freedom of the gravitational theory. The optical critical curve instead follows from null rays approaching the unstable photon region. In general relativity (GR), these descriptions are related in the eikonal regime. Additional scalar and vector fields in modified gravity need not obey the same relation. Predicting a dynamical shadow therefore requires a calculation of both the field perturbations and their effect on null propagation within the theory.

Scalar--tensor--vector gravity (STVG), also called modified gravity (MOG), contains the metric, a vector field, and scalar fields including the gravitational coupling $G$ and vector mass scale $\mu$ \cite{Moffat:2005si}. Its standard nonrotating black-hole solution uses a reduced compact-object regime with background coupling $G=G_{\mathrm N}(1+\alpha)$ and a gravitational vector charge fixed by the source mass \cite{Moffat:2014aja}. The Schwarzschild-MOG metric has the algebraic form of the Reissner--Nordstr\"om (RN) metric. Static calculations have examined its horizons, photon sphere, shadow size, and accretion-dependent ring structure \cite{Moffat:2015kva,Hu:2022lek}. Perturbative studies have identified tensor, vector-induced, and scalar polarizations in weak-field STVG \cite{Liu:2019cxm} and calculated gravito-vector quasinormal modes (QNMs) and black-hole ringdown waveforms \cite{Manfredi:2017xcv,Liu:2023uft}.

Time-dependent photon structures require global null propagation rather than an instantaneous change to a static photon-sphere radius \cite{Koga:2022dsu}. Ray-tracing studies have calculated how prescribed spacetime perturbations deform a Schwarzschild shadow \cite{Wang:2019skw}. Perturbative methods also relate static and QNM-driven metric perturbations to shadow displacements \cite{Kobialko:2024zhc,Pantig:2025eqe}. These methods describe the image-plane response to a known metric perturbation, but they do not derive the scalar and gravito-vector perturbations of a Schwarzschild-MOG black hole from the STVG field equations or test whether a dynamical scalar perturbation is optically nontrivial for null critical rays.

A phenomenological MOG model proposed a scalar-driven breathing shadow followed by a delayed vector-driven translation \cite{Lobos:2026jyp}. A perturbative treatment of the field equations can test this sequence. The nonzero background vector field couples metric and vector perturbations at first order. The scalar monopole must satisfy its own wave equation rather than take a large-$\ell$ photon-sphere frequency. Neglecting the background correction from a finite vector mass while retaining its propagation effects requires a controlled hierarchy of scales. The image displacement must also follow from null propagation and an observer frame, not from individual gauge-dependent metric components.

This paper establishes a scalar critical-curve invariance result within a reduced strong-field MOG sector. The local calculation retains a dynamical fluctuation of $G$, fixes the vector coupling, and works at leading order in $\mu\bar G M\ll1$. A separate far-zone calculation accounts for finite-range vector propagation. The analysis presents the canonical $\ell\ge2$ RN master system and treats the scalar monopole and exceptional axial dipole separately. It does not develop the polar dipole reconstruction. For the axial dipole, a gauge-invariant constraint and an explicit horizon-regular gauge give a first-order metric reconstruction and a controlled local low-frequency limit. The null Hamiltonian and a specified observer-frame identification then establish the scalar optical result. A stationary charged rotating perturbation tests the metric-to-optics normalization. The analysis does not calculate the finite-frequency scattering-to-image response or introduce a new MOG mode family beyond the inherited RN spectrum.

Section~\ref{sec:background} presents the reduced action and the black-hole background. Section~\ref{sec:linear_perturbations} derives the first-order master equations and the gauge-fixed axial-dipole reconstruction. Section~\ref{sec:qnm} gives the spectral boundary conditions, stability analysis, eikonal limits, and far-zone vector propagation. Section~\ref{sec:photon_ring} proves scalar critical-curve invariance, derives the local low-frequency scattering limit, and gives a stationary optical benchmark. Section~\ref{sec:consistency} checks the GR and decoupling limits, dimensions, and scale separation. Section~\ref{sec:conclusion} summarizes the results, their contribution to the perturbation and optical analysis, and the remaining limits.

\section{STVG action and Schwarzschild-MOG background}
\label{sec:background}

The original STVG theory treats the gravitational coupling $G$, vector coupling $\omega$, and inverse-range parameter $\mu$ as dynamical scalar fields, alongside the metric $g_{\mu\nu}$ and vector field $\phi_\mu$ \cite{Moffat:2005si}. The standard Schwarzschild-MOG solution instead assumes constant background $G$, fixed $\omega$, and a vector mass negligible on black-hole length scales \cite{Moffat:2014aja}. The reduced action below retains a dynamical fluctuation of $G$ and admits this background. It describes the effective strong-field sector considered here, not the full STVG theory or a truncation valid for arbitrary configurations. The local calculation fixes $\omega$, treats $\mu$ as an external inverse-range scale, and expands in $\mu\bar G M\ll1$. A finite-$\mu$ strong-field Proca treatment, including the longitudinal vector mode and the corresponding background correction, lies outside this sector.

The metric signature is $(-,+,+,+)$, and units satisfy $c=\hbar=1$. The curvature convention is
\begin{equation}
[\nabla_\mu,\nabla_\nu]V_\lambda
=
R^\rho{}_{\lambda\mu\nu}V_\rho ,
\label{eq:riemann_convention_sec2}
\end{equation}
with Ricci tensor
\begin{equation}
R_{\mu\nu}=R^\rho{}_{\mu\nu\rho}
\label{eq:ricci_convention}
\end{equation}
and Ricci scalar $R=g^{\mu\nu}R_{\mu\nu}$. The reduced action is
\begin{multline}
S=\int d^4x\,\sqrt{-g}\Bigg[
\frac{R}{16\pi G}-\frac{1}{16\pi}B_{\mu\nu}B^{\mu\nu}
-\frac{\mu^2}{8\pi}\phi_\mu\phi^\mu
\\
-\frac{1}{2G^3}\nabla_\mu G\nabla^\mu G
-\frac{V_G(G)}{G^3}\Bigg]+S_M,
\label{eq:reduced_mog_action}
\end{multline}
where
\begin{equation}
B_{\mu\nu}
=
2\nabla_{[\mu}\phi_{\nu]}
=
\partial_\mu\phi_\nu-\partial_\nu\phi_\mu .
\label{eq:B_definition_sec2}
\end{equation}
The vector kinetic term is normalized to give the standard Maxwell stress tensor in the massless limit. The Proca term in Eq.~\eqref{eq:reduced_mog_action} has the sign appropriate to the mostly-plus signature. The scalar potential $V_G$ is unspecified apart from the vacuum conditions below. Matter is absent from the black-hole exterior, but $S_M$ is retained to define the vector current.

Variation with respect to the inverse metric gives
\begin{equation}
\begin{split}
&G_{\mu\nu}+G\left(g_{\mu\nu}\Box-\nabla_\mu\nabla_\nu\right)G^{-1}
\\
&\qquad=8\pi G\left(T^{(\phi)}_{\mu\nu}+T^{(G)}_{\mu\nu}+T^{(M)}_{\mu\nu}\right),
\end{split}
\label{eq:metric_field_eq_sec2}
\end{equation}
where $\Box=\nabla_\rho\nabla^\rho$ and
\begin{equation}
\begin{split}
T^{(\phi)}_{\mu\nu}
=
\frac{1}{4\pi}
\Bigg[
&
B_{\mu\rho}B_\nu{}^\rho
-\frac{1}{4}g_{\mu\nu}B_{\rho\sigma}B^{\rho\sigma}
\\
&
+\mu^2
\left(
\phi_\mu\phi_\nu
-\frac{1}{2}g_{\mu\nu}\phi_\rho\phi^\rho
\right)
\Bigg],
\end{split}
\label{eq:vector_stress_sec2}
\end{equation}
while
\begin{equation}
\begin{split}
T^{(G)}_{\mu\nu}=\frac{1}{G^3}\Bigl[
&\nabla_\mu G\nabla_\nu G
-\frac{1}{2}g_{\mu\nu}\nabla_\rho G\nabla^\rho G
\\
&-g_{\mu\nu}V_G(G)\Bigr].
\end{split}
\label{eq:scalar_stress_sec2}
\end{equation}
The matter current is defined by
\begin{equation}
J^\mu
=
-\frac{1}{\sqrt{-g}}
\frac{\delta S_M}{\delta\phi_\mu},
\label{eq:current_definition_sec2}
\end{equation}
and variation with respect to $\phi_\mu$ yields
\begin{equation}
\nabla_\nu B^{\nu\mu}
-
\mu^2\phi^\mu
=
4\pi J^\mu .
\label{eq:vector_field_eq_sec2}
\end{equation}
The $G$ variation gives
\begin{equation}
\Box G
-
\frac{3}{2G}\nabla_\rho G\nabla^\rho G
-
\frac{G}{16\pi}R
+
\frac{3V_G(G)}{G}
-
V_G'(G)
=
0 .
\label{eq:G_field_eq_sec2}
\end{equation}
Equations \eqref{eq:metric_field_eq_sec2}--\eqref{eq:G_field_eq_sec2} define the local field theory for the perturbation analysis. No field equation for $\mu$ is imposed because $\mu$ is not varied in this reduced sector.

The zeroth-order black-hole geometry has coupling
\begin{equation}
\bar G=G_{\mathrm N}(1+\alpha),
\label{eq:Gbar_alpha}
\end{equation}
where $\alpha\geq0$, and satisfies the vacuum conditions $V_G(\bar G)=V_G'(\bar G)=0$. The local compact-object background retains leading-order terms in the dimensionless parameter $\epsilon_\mu\equiv\mu\bar G M\ll1$.
Setting $\mu=0$ in the background equations gives the zeroth-order near-zone solution. This is the approximation used in the standard Schwarzschild-MOG construction \cite{Moffat:2014aja}. A separate far-zone expansion describes finite-range propagation. The local equations neglect both the vector mass and its backreaction at the retained order. The ratio of the Proca mass term to the vector kinetic term scales as
\begin{equation}
\frac{\mu^2\phi_\rho\phi^\rho}
{B_{\rho\sigma}B^{\rho\sigma}}
=\mathcal{O}\!\left((\mu r)^2\right),
\label{eq:proca_near_zone_scaling}
\end{equation}
so corrections to the photon-region background are of order $\epsilon_\mu^2$ when $r=\mathcal{O}(\bar G M)$.

In the vacuum near zone, $J^\mu=0$, $T^{(M)}_{\mu\nu}=0$, and $\nabla_\mu\bar G=0$. The scalar stress tensor vanishes, and the massless vector stress tensor is traceless. The trace of Eq.~\eqref{eq:metric_field_eq_sec2} then gives $\bar R=0$. The constant background \eqref{eq:Gbar_alpha} and vacuum conditions $V_G(\bar G)=V_G'(\bar G)=0$ satisfy Eq.~\eqref{eq:G_field_eq_sec2}. The constant-$G$ Schwarzschild-MOG geometry is therefore an exact solution of the reduced near-zone theory.

The static, spherically symmetric line element is
\begin{equation}
d\bar s^2
=
-f(r)\,dt^2
+
\frac{dr^2}{f(r)}
+
r^2
\left(
d\theta^2
+
\sin^2\theta\,d\varphi^2
\right)
\label{eq:mog_background_metric_sec2}
\end{equation}
with electric-type gravitational vector potential
\begin{equation}
\bar\phi_\mu dx^\mu
=
\frac{Q_g}{r}\,dt .
\label{eq:background_vector_sec2}
\end{equation}
The corresponding nonzero field-strength component is
\begin{equation}
\bar B_{tr}
=
\frac{Q_g}{r^2}.
\label{eq:background_Btr_sec2}
\end{equation}
Equation \eqref{eq:vector_field_eq_sec2} with $\mu=0$ reduces to
\begin{equation}
\partial_r
\left(
r^2\bar B^{rt}
\right)
=
0 ,
\label{eq:background_vector_integral_sec2}
\end{equation}
which confirms that $Q_g$ is the conserved gravitational vector charge. The mixed components of the vector stress tensor are
\begin{equation}
T^{(\phi)\mu}{}_{\nu}
=
\frac{Q_g^2}{8\pi r^4}
\operatorname{diag}(-1,-1,1,1).
\label{eq:background_stress_sec2}
\end{equation}
For the metric \eqref{eq:mog_background_metric_sec2}, the $t$-$t$ field equation becomes
\begin{equation}
r f'(r)+f(r)-1
=
-\frac{\bar G Q_g^2}{r^2}.
\label{eq:background_f_ode_sec2}
\end{equation}
Integration gives
\begin{equation}
f(r)
=
1
-
\frac{2\bar G M}{r}
+
\frac{\bar G Q_g^2}{r^2},
\label{eq:background_f_charge_sec2}
\end{equation}
where $M$ is the asymptotic mass parameter. The MOG source-charge relation is
\begin{equation}
Q_g
=
\sqrt{\alpha G_{\mathrm N}}\,M ,
\label{eq:mog_charge_relation_sec2}
\end{equation}
for the stationary background. In the perturbation problem, $Q_g$ remains a conserved background integration constant. A local fluctuation of $G$ does not impose an algebraic variation $\delta Q_g$. Charge-changing monopoles describe different stationary members of the background family and are excluded by fixing the asymptotic charges. Substituting Eq.~\eqref{eq:mog_charge_relation_sec2} into Eq.~\eqref{eq:background_f_charge_sec2} gives
\begin{equation}
f(r)
=
1
-
\frac{2G_{\mathrm N}(1+\alpha)M}{r}
+
\frac{\alpha(1+\alpha)G_{\mathrm N}^2M^2}{r^2}.
\label{eq:mog_lapse_sec2}
\end{equation}
The metric has the RN form, but $Q_g$ is a gravitational vector charge tied to $M$ through Eq.~\eqref{eq:mog_charge_relation_sec2}, not an independent electromagnetic charge \cite{Moffat:2014aja}.

The roots of $f(r)=0$ are
\begin{equation}
r_\pm
=
G_{\mathrm N}M
\left[
1+\alpha
\pm
\sqrt{1+\alpha}
\right].
\label{eq:mog_horizons_sec2}
\end{equation}
For $\alpha>0$, $r_+$ is the event horizon and $r_-$ is the inner horizon. The limit $\alpha\rightarrow0$ gives $r_+\rightarrow2G_{\mathrm N}M$ and recovers the Schwarzschild exterior.

The unperturbed critical curve follows from null geodesics of Eq.~\eqref{eq:mog_background_metric_sec2}. Spherical symmetry allows restriction to the equatorial plane, where the Killing integrals are $E=f(r)\dot t$ and $L=r^2\dot\varphi$. An overdot denotes differentiation with respect to an affine parameter. The null condition gives
\begin{equation}
\dot r^2
+
\frac{L^2 f(r)}{r^2}
=
E^2 .
\label{eq:null_radial_eq_sec2}
\end{equation}
A circular null orbit satisfies
\begin{equation}
r f'(r)-2f(r)=0 .
\label{eq:photon_sphere_condition_sec2}
\end{equation}
The outer solution is
\begin{equation}
r_{\mathrm{ph}}
=
\frac{G_{\mathrm N}M}{2}
\left[
3(1+\alpha)
+
\sqrt{(1+\alpha)(9+\alpha)}
\right].
\label{eq:photon_sphere_radius_sec2}
\end{equation}
For $\alpha\geq0$, this orbit is unstable because $d^2(f/r^2)/dr^2<0$ at $r=r_{\mathrm{ph}}$. The critical impact parameter is
\begin{equation}
b_c^2
=
\frac{r_{\mathrm{ph}}^2}{f(r_{\mathrm{ph}})}
=
\frac{2r_{\mathrm{ph}}^3}
{r_{\mathrm{ph}}-\bar G M}.
\label{eq:critical_impact_sec2}
\end{equation}
For a static observer at radius $r_o$ in the asymptotically flat region, the local angular radius $\vartheta_c$ of the critical curve satisfies
\begin{equation}
\sin\vartheta_c
=
\frac{b_c\sqrt{f(r_o)}}{r_o}.
\label{eq:observer_angle_sec2}
\end{equation}
At asymptotic distance, the corresponding screen radius is $R_0=b_c$. In the general-relativistic limit $\alpha\to0$, the photon-sphere radius and critical impact parameter approach $r_{\mathrm{ph}}\to3G_{\mathrm N}M$ and $b_c\to3\sqrt{3}\,G_{\mathrm N}M$.

Equations \eqref{eq:mog_background_metric_sec2}--\eqref{eq:critical_impact_sec2} define the zeroth-order near-zone geometry and optical critical curve. Perturbing the fields in Eq.~\eqref{eq:reduced_mog_action} about this background retains the first-order coupling between vector and metric perturbations due to the nonzero field strength \eqref{eq:background_Btr_sec2}.

\section{Gauge-invariant perturbations and exact master reduction}
\label{sec:linear_perturbations}

The perturbation expansion of the reduced near-zone theory is first order in a dimensionless parameter $\epsilon$ and leading order in $\mu\bar G M$. The local vector field is therefore massless and retains its Abelian gauge symmetry. The fields are
\begin{align}
g_{\mu\nu}&=\bar g_{\mu\nu}+\epsilon h_{\mu\nu},
\label{eq:prd_field_perturbations}
\\
\phi_\mu&=\bar\phi_\mu+\epsilon a_\mu,
\label{eq:prd_vector_perturbation}
\\
G&=\bar G(1+\epsilon\psi),
\label{eq:prd_scalar_perturbation}
\end{align}
with $\bar G=G_{\mathrm N}(1+\alpha)$. The metric trace is $h=\bar g^{\mu\nu}h_{\mu\nu}$, and the perturbed field strength is $b_{\mu\nu}=2\bar\nabla_{[\mu}a_{\nu]}$.
Because $\bar\nabla_\mu\bar G=0$ and $V_G(\bar G)=V_G'(\bar G)=0$, the explicit scalar stress tensor begins at second order. The linear metric equation is
\begin{equation}
\begin{split}
&\delta G_{\mu\nu}+\bar\nabla_\mu\bar\nabla_\nu\psi-\bar g_{\mu\nu}\bar\Box\psi
\\
&\qquad=8\pi\bar G\left(\delta T^{(\phi)}_{\mu\nu}+\psi\bar T^{(\phi)}_{\mu\nu}\right),
\end{split}
\label{eq:prd_linear_metric}
\end{equation}
where
\begin{equation}
\begin{split}
4\pi\delta T^{(\phi)}_{\mu\nu}={}&
b_{\mu\rho}\bar B_\nu{}^\rho+\bar B_{\mu\rho}b_\nu{}^\rho
-\bar B_{\mu\rho}h^{\rho\sigma}\bar B_{\nu\sigma}
\\
&-\frac14 h_{\mu\nu}\bar B_{\rho\sigma}\bar B^{\rho\sigma}
-\frac14\bar g_{\mu\nu}\delta(B^2),
\end{split}
\label{eq:prd_delta_T}
\end{equation}
and
\begin{equation}
\delta(B^2)=2\bar B^{\rho\sigma}b_{\rho\sigma}
-2h^{\rho\sigma}\bar B_{\rho\lambda}\bar B_\sigma{}^\lambda.
\label{eq:prd_delta_B2}
\end{equation}
The linear vector equation is
\begin{equation}
\bar\nabla_\nu\left[
b^{\nu\mu}-h^\nu{}_\rho\bar B^{\rho\mu}
-h^\mu{}_\rho\bar B^{\nu\rho}
+\frac12h\bar B^{\nu\mu}
\right]=0.
\label{eq:prd_linear_vector}
\end{equation}
The scalar equation obtained from the $G$ variation is
\begin{equation}
\bar\Box\psi-\frac{\delta R}{16\pi}-V_G''(\bar G)\psi=0.
\label{eq:prd_scalar_pretrace}
\end{equation}
The terms proportional to $\bar B_{\mu\nu}b_{\rho\sigma}$ in Eqs.~\eqref{eq:prd_linear_metric}--\eqref{eq:prd_scalar_pretrace} couple the metric and vector perturbations at first order.

The scalar coupling can be separated by defining
\begin{equation}
\widehat h_{\mu\nu}=h_{\mu\nu}-\psi\bar g_{\mu\nu}.
\label{eq:prd_hhat_def}
\end{equation}
For a conformal perturbation $\delta g_{\mu\nu}=\psi\bar g_{\mu\nu}$ in four dimensions with $\bar R=0$,
\begin{equation}
\delta G_{\mu\nu}[\psi\bar g]
=-\bar\nabla_\mu\bar\nabla_\nu\psi
+\bar g_{\mu\nu}\bar\Box\psi.
\label{eq:prd_conformal_Einstein}
\end{equation}
The Maxwell-type stress tensor obeys the first-order conformal identity
\begin{equation}
\delta T^{(\phi)}_{\mu\nu}[\psi\bar g,0]
=-\psi\bar T^{(\phi)}_{\mu\nu},
\label{eq:prd_conformal_Maxwell}
\end{equation}
and the conformal terms in the bracket of Eq.~\eqref{eq:prd_linear_vector} cancel because $h^\mu{}_\nu=\psi\delta^\mu{}_\nu$ and $h=4\psi$. Substitution of Eq.~\eqref{eq:prd_hhat_def} gives
\begin{equation}
\delta G_{\mu\nu}[\widehat h]
=8\pi\bar G\,\delta T^{(\phi)}_{\mu\nu}[\widehat h,a],
\label{eq:prd_decoupled_metric}
\end{equation}
\begin{equation}
\bar\nabla_\nu\left[
b^{\nu\mu}-\widehat h^\nu{}_\rho\bar B^{\rho\mu}
-\widehat h^\mu{}_\rho\bar B^{\nu\rho}
+\frac12\widehat h\bar B^{\nu\mu}
\right]=0.
\label{eq:prd_decoupled_vector}
\end{equation}
The trace of Eq.~\eqref{eq:prd_decoupled_metric} gives $\delta R[\widehat h]=0$ because the full Maxwell stress tensor is traceless. Since $\delta R[\psi\bar g]=-3\bar\Box\psi$, Eq.~\eqref{eq:prd_scalar_pretrace} becomes
\begin{equation}
\left(\bar\Box-m_G^2\right)\psi=0,
\label{eq:prd_scalar_KG}
\end{equation}
where the effective scalar mass is
\begin{equation}
m_G^2=\frac{16\pi}{16\pi+3}V_G''(\bar G).
\label{eq:prd_scalar_mass}
\end{equation}
The scalar does not source the gravito-vector system at first order in this reduced theory. Its contribution to the original metric variable is the conformal term in $h_{\mu\nu}=\widehat h_{\mu\nu}+\psi\bar g_{\mu\nu}$.

Under an infinitesimal coordinate transformation generated by $\xi^\mu$,
\begin{equation}
\widehat h_{\mu\nu}\rightarrow\widehat h_{\mu\nu}-2\bar\nabla_{(\mu}\xi_{\nu)},
\label{eq:prd_gauge_hhat}
\end{equation}
while
\begin{equation}
a_\mu\rightarrow a_\mu-\mathcal L_\xi\bar\phi_\mu+\bar\nabla_\mu\Lambda.
\label{eq:prd_gauge_a}
\end{equation}
The scalar is gauge invariant, $\psi\to\psi$, because $\bar G$ is constant. The Moncrief and Gerlach--Sengupta constructions give gauge-invariant spherical perturbation variables from $\widehat h_{\mu\nu}$ and $a_\mu$ \cite{Moncrief:1974gw,Moncrief:1974ng,Moncrief:1975sb, Gerlach:1979rw,Martel:2005ir}. The indices $a,b\in\{t,r\}$ label the orbit space, and $A,B$ label the sphere. Scalar harmonics obey $D^AD_AY_{\ell m}=-\ell(\ell+1)Y_{\ell m}$. Even vector harmonics are $Y_A=D_AY$, and odd vector harmonics are $S_A=\epsilon_A{}^BD_BY$. The scalar perturbation has only even parity and takes the form
\begin{equation}
\psi=\sum_{\ell m}\frac{\Phi_{\ell m}(t,r)}{r}Y_{\ell m}.
\label{eq:prd_scalar_decomp}
\end{equation}
It satisfies
\begin{equation}
\left[-\partial_t^2+\partial_{r_*}^2-V_{s,\ell}(r)\right]\Phi_{\ell m}=0,
\label{eq:prd_scalar_master}
\end{equation}
with
\begin{equation}
V_{s,\ell}=f\left[\frac{\ell(\ell+1)}{r^2}+\frac{f'}{r}+m_G^2\right].
\label{eq:prd_scalar_potential}
\end{equation}
This equation also governs the monopole $\ell=0$, without an eikonal approximation.

The field rescaling $\mathcal A_\mu=\sqrt{\bar G}\,\phi_\mu$ puts the remaining gravito-vector equations in the Einstein--Maxwell RN form. The effective mass and charge parameters are
\begin{align}
\mathcal M&=\bar G M,
\label{eq:prd_RN_defs}
\\
\mathcal Q&=-\sqrt{\bar G}\,Q_g,
\qquad \mathcal Q^2=\bar GQ_g^2.
\label{eq:prd_RN_charge}
\end{align}
Thus $\bar{\mathcal A}_t=-\mathcal Q/r$, matching the signed-charge convention of Ref.~\cite{Burton:2020wnj}, even though the original MOG potential is $+Q_g/r$. The canonically rescaled vector perturbation is $\mathfrak a_\mu=\sqrt{\bar G}\,a_\mu$. Below, $a_d$ denotes a harmonic amplitude of $\mathfrak a_\mu$, not of the unrescaled $a_\mu$. Background quantities and master potentials depend on $\mathcal Q^2$, whereas the metric response to a specified canonical vector amplitude is odd in the signed $\mathcal Q$.
The radial function
\begin{equation}
\Delta=r^2-2\mathcal M r+\mathcal Q^2
\label{eq:prd_Delta}
\end{equation}
gives $f=\Delta/r^2$. For $\ell\ge2$, the angular parameter is $n=(\ell-1)(\ell+2)/2$, and the characteristic coefficients are
\begin{equation}
p_{1,2}=3\mathcal M\pm\sqrt{9\mathcal M^2+8n\mathcal Q^2}.
\label{eq:prd_p12}
\end{equation}
The canonical axial master functions $Z_i^{(-)}$ satisfy
\begin{equation}
\left[-\partial_t^2+\partial_{r_*}^2-V_i^{(-)}\right]Z_i^{(-)}=0,
\label{eq:prd_axial_master}
\end{equation}
where $i,j\in\{1,2\}$, $i\ne j$, and
\begin{equation}
V_i^{(-)}=\frac{\Delta}{r^5}\left[2(n+1)r-p_j+\frac{4\mathcal Q^2}{r}\right].
\label{eq:prd_axial_potential}
\end{equation}
Diagonalization of the polar system uses the functions
\begin{equation}
\varpi=nr+3\mathcal M-\frac{2\mathcal Q^2}{r},
\label{eq:prd_varpi}
\end{equation}
\begin{equation}
W=\frac{\Delta(2nr+3\mathcal M)}{r\varpi^2}+\frac{nr+\mathcal M}{\varpi},
\label{eq:prd_W}
\end{equation}
\begin{equation}
U=(2nr+3\mathcal M)W+\varpi-nr-\mathcal M-\frac{2n\Delta}{\varpi}.
\label{eq:prd_U}
\end{equation}
The two polar master functions obey
\begin{equation}
\left[-\partial_t^2+\partial_{r_*}^2-V_i^{(+)}\right]Z_i^{(+)}=0,
\label{eq:prd_polar_master}
\end{equation}
with
\begin{align}
V_1^{(+)}&=\frac{\Delta}{r^5}\left[U+\frac{p_1-p_2}{2}W\right],
\label{eq:prd_polar_potentials}
\\
V_2^{(+)}&=\frac{\Delta}{r^5}\left[U-\frac{p_1-p_2}{2}W\right].
\label{eq:prd_polar_potential_2}
\end{align}
These are the RN polar potentials in the asymptotically flat limit \cite{Chandrasekhar:1979iz,Liu:2025reu}, with normalization fixed by the neutral limit. As $\mathcal Q\rightarrow0$, $V_1^{(-)}$ and $V_1^{(+)}$ both reduce to the electromagnetic potential $f\ell(\ell+1)/r^2$, $V_2^{(-)}$ reduces to the Regge--Wheeler potential, and $V_2^{(+)}$ reduces to the Zerilli potential \cite{Regge:1957td,Zerilli:1970se}.

The even-parity master equations for $\ell\ge2$ take the matrix form
\begin{equation}
\left[-\partial_t^2+\partial_{r_*}^2-\mathbf V_\ell^{(+)}(r)\right]
\mathbf\Psi_{\ell m}^{(+)}=0,
\label{eq:prd_full_even_matrix}
\end{equation}
with
\begin{align}
\mathbf\Psi_{\ell m}^{(+)}&=
\begin{pmatrix}Z_1^{(+)}\\[2pt] Z_2^{(+)}\\[2pt] \Phi\end{pmatrix},
\label{eq:prd_full_even_vector}
\\
\mathbf V_\ell^{(+)}&=
\operatorname{diag}\left(V_1^{(+)},V_2^{(+)},V_{s,\ell}\right).
\label{eq:prd_full_even_diag}
\end{align}
The physical metric and vector variables are linear combinations of $Z_1$ and $Z_2$. Gravito-vector mixing therefore remains present even though the canonical RN basis diagonalizes the radial operator. The off-diagonal entries in the scalar row and column vanish.

The low multipoles require separate constrained reductions. For $\ell=0$, the RN gravito-vector sector contains only stationary changes in the background mass and charge. Fixing the Arnowitt--Deser--Misner (ADM) mass and MOG charge excludes these variations. The remaining dynamical monopole is
\begin{equation}
h_{\mu\nu}^{(s,0)}=\psi_{00}\bar g_{\mu\nu},
\label{eq:prd_scalar_monopole_metric}
\end{equation}
with $a_\mu^{(s,0)}=0$ and $\Phi_{00}$ governed by Eqs.~\eqref{eq:prd_scalar_master} and \eqref{eq:prd_scalar_potential}. For $\ell=1$, the usual $\ell\ge2$ Moncrief transformation is singular because the odd gravitational tensor harmonic vanishes and there is no independent radiative gravitational dipole. This exceptional sector has been treated separately in the Reissner--Nordstr\"om literature \cite{Burko:1999jy,Burton:2020wnj}. The regular vector-harmonic variables of Ref.~\cite{Burton:2020wnj} make both the Einstein--Maxwell constraints and global angular regularity explicit. The odd vector harmonic is
\begin{equation}
X_A^{\ell m}=-\epsilon_A{}^B D_B Y_{\ell m},
\label{eq:prd_odd_vector_harmonic_dipole}
\end{equation}
where $D_A$ and $\epsilon_{AB}$ are the covariant derivative and volume form of the unit two-sphere. The axial $\ell=1$ perturbations are written as
\begin{align}
\mathfrak a_A^{(d)}&=a_d(r,\omega)X_A^{1m}e^{-i\omega t},
\label{eq:prd_dipole_vector_regular}
\\
h_{aB}^{(d)}&=h_a(r,\omega)X_B^{1m}e^{-i\omega t},
\qquad
h_{AB}^{(d)}=0.
\label{eq:prd_dipole_metric_regular}
\end{align}
The last equality is not a gauge assumption. The odd tensor harmonic $X_{AB}^{1m}=D_{(A}X_{B)}^{1m}$ vanishes identically for $\ell=1$. For the axis-aligned mode $Y_{10}=N_1\cos\theta$, with $N_1=\sqrt{3/(4\pi)}$, Eq.~\eqref{eq:prd_odd_vector_harmonic_dipole} gives
\begin{equation}
X_\theta^{10}=0,
\qquad
X_\varphi^{10}=-N_1\sin^2\theta.
\label{eq:prd_dipole_axis_regular}
\end{equation}
Hence $h_{t\varphi}$ and $h_{r\varphi}$ vanish as $\sin^2\theta$ at both symmetry axes, and there is no $h_{\theta\varphi}$ component. The dipole reconstruction is therefore globally regular rather than merely regular on the equatorial plane.

Before diagonalization, the source-free $\ell=1$ odd Einstein--Maxwell equations retain their constrained form. In the normalization used here, the angular Maxwell equation is
\begin{equation}
\begin{split}
0={}&f\frac{d^2a_d}{dr^2}+f'\frac{da_d}{dr}
+\left(\frac{\omega^2}{f}-\frac{2}{r^2}\right)a_d
\\
&+\frac{\mathcal Q}{r^2}\frac{dh_t}{dr}
-\frac{2\mathcal Q}{r^3}h_t
+\frac{i\omega\mathcal Q}{r^2}h_r,
\end{split}
\label{eq:prd_dipole_maxwell_constrained}
\end{equation}
while the gauge-invariant metric combination is
\begin{equation}
\mathcal K_d=h_t'-\frac{2h_t}{r}+i\omega h_r.
\label{eq:prd_dipole_invariant}
\end{equation}
An odd coordinate transformation generated by $\xi_A=\zeta(r)X_A^{1m}e^{-i\omega t}$ changes the metric amplitudes according to
\begin{align}
h_t&\longrightarrow h_t+i\omega\zeta,
\\
h_r&\longrightarrow h_r-\zeta'+2\zeta/r,
\label{eq:prd_dipole_gauge}
\end{align}
and leaves both $\mathcal K_d$ and $a_d$ unchanged. In the source-free exterior the two nonvanishing odd Einstein equations are equivalent to
\begin{equation}
\partial_r\mathfrak j_d=0,
\qquad \partial_t\mathfrak j_d=0,
\qquad
\mathfrak j_d=r^2\mathcal K_d+4\mathcal Q a_d.
\label{eq:prd_dipole_angular_momentum}
\end{equation}
Here $\mathfrak j_d$ is a stationary angular-momentum integration constant, with its normalization checked below. It vanishes for a nonzero Fourier frequency. Fixing zero angular momentum also sets it to zero in the static sector. Thus
\begin{equation}
h_t'-\frac{2h_t}{r}+i\omega h_r
=-\frac{4\mathcal Q}{r^2}a_d
\label{eq:prd_dipole_constraint_1}
\end{equation}
in the scattering problem. For example, in the exterior gauge $h_r=0$, the radial derivative of Eq.~\eqref{eq:prd_dipole_angular_momentum} gives $h_t''-2h_t/r^2+4\mathcal Q a_d'/r^2=0$, and its time derivative gives the remaining odd Einstein equation. These are the source-free equations of Appendix~C.2 of Ref.~\cite{Burton:2020wnj}.

There is no additional tensor-harmonic Einstein equation at $\ell=1$. In particular, $(fh_r)'+i\omega h_t/f=0$, which occurs in the higher-multipole reconstruction, is not a dipole constraint because its variation under Eq.~\eqref{eq:prd_dipole_gauge} is generally nonzero. Imposing it would be a separate gauge choice requiring its own residual-gauge and boundary analysis. The present reconstruction does not impose this additional gauge condition.

Substituting Eq.~\eqref{eq:prd_dipole_constraint_1} into Eq.~\eqref{eq:prd_dipole_maxwell_constrained} cancels the metric amplitudes exactly and gives
\begin{equation}
\left[
\frac{d^2}{dr_*^2}+\omega^2-V_{v1}(r)
\right]a_d(r,\omega)=0,
\label{eq:prd_dipole_vector_master}
\end{equation}
with
\begin{equation}
V_{v1}(r)=f(r)\left(
\frac{2}{r^2}+\frac{4\mathcal Q^2}{r^4}
\right).
\label{eq:prd_dipole_vector_potential}
\end{equation}
This is the source-free electromagnetic-type dipole potential of the regular RN formulation \cite{Burton:2020wnj}. It is positive outside a subextremal horizon. If a stationary $\mathfrak j_d\ne0$ is retained, the right-hand side of Eq.~\eqref{eq:prd_dipole_vector_master} is instead $-f\mathcal Q\mathfrak j_d/r^4$. That sector is used only for the stationary benchmark in Sec.~\ref{sec:photon_ring}.

Ingoing coordinates $v=t+r_*$ allow reconstruction of the zero-angular-momentum metric without division by frequency. The corresponding amplitudes are
\begin{equation}
\widetilde a_d=e^{i\omega r_*}a_d,
\qquad \widetilde h_v=e^{i\omega r_*}h_t.
\label{eq:prd_dipole_ingoing_amplitudes}
\end{equation}
The reconstruction uses the ingoing radial gauge
\begin{equation}
\widetilde h_r=e^{i\omega r_*}(h_r-h_t/f)=0.
\label{eq:prd_dipole_radial_gauge}
\end{equation}
This is an explicit coordinate choice, not an Einstein equation. A regular odd gauge generator changes $\widetilde h_r$ by $-\widetilde\zeta'+2\widetilde\zeta/r$, so a first-order radial integration gives this gauge. Its remaining freedom is $\widetilde\zeta=c(v)r^2$, an angular rotation. Fixing the angular frame at infinity removes its time-dependent part. A fixed screen orientation removes the constant rotation.

At fixed $v$, Eq.~\eqref{eq:prd_dipole_constraint_1} becomes
\begin{equation}
\frac{d}{dr}\left(\frac{\widetilde h_v}{r^2}\right)
=-\frac{4\mathcal Q}{r^4}\widetilde a_d.
\label{eq:prd_dipole_radial_constraint}
\end{equation}
With $\widetilde h_v/r^2\to0$ at infinity, its unique solution is
\begin{equation}
\widetilde h_v(r,\omega)
=4\mathcal Q r^2\int_r^\infty
\frac{\widetilde a_d(s,\omega)}{s^4}\,ds.
\label{eq:prd_dipole_metric_integral}
\end{equation}
For a bounded real-frequency scattering mode the integral converges. An ingoing vector mode has finite $\widetilde a_d$ at the future horizon, and Eq.~\eqref{eq:prd_dipole_metric_integral} gives finite $\widetilde h_v$ there. The original Schwarzschild-coordinate amplitudes are $h_t=e^{-i\omega r_*}\widetilde h_v$ and $h_r=h_t/f$. Their coordinate behavior does not imply a singular ingoing metric. No independent gravitational radiation condition is needed for this constrained dipole. The vector excitation satisfies the causal boundary conditions, while the radial integral fixes the metric representative and angular frame. When $\mathcal Q\to0$ at fixed zero $\mathfrak j_d$, the reconstructed metric vanishes and the vector equation reduces to the Maxwell dipole on Schwarzschild.

\section{Spectrum, stability, and propagation hierarchy}
\label{sec:qnm}

The spectral problem uses time dependence $e^{-i\omega t}$ and tortoise coordinate $dr_*/dr=f^{-1}$. The horizons are
\begin{equation}
r_\pm=\mathcal M\pm\sqrt{\mathcal M^2-\mathcal Q^2}.
\label{eq:prd_horizons_eff}
\end{equation}
The MOG charge relation gives $\mathcal Q^2/\mathcal M^2=\alpha/(1+\alpha)<1$ for every finite $\alpha\ge0$, so the local black-hole family is subextremal throughout this range. A massless gravito-vector master function obeys the quasinormal boundary conditions
\begin{align}
Z&=A_{\mathrm H}e^{-i\omega r_*}\left[1+o(1)\right]
\quad(r_*\to-\infty),
\label{eq:prd_qnm_bc}
\\
Z&=A_{\infty}e^{+i\omega r_*}\left[1+o(1)\right]
\quad(r_*\to+\infty).
\label{eq:prd_qnm_bc_outgoing}
\end{align}
Here $A_{\mathrm H}$ and $A_{\infty}$ are nonzero normalization constants.
For solutions $Z^{\rm in}$ and $Z^{\rm up}$ satisfying the ingoing and outgoing conditions, respectively, the QNM frequencies are the zeros of the Wronskian,
\begin{equation}
\mathcal W_I(\omega)=W[Z_I^{\rm in},Z_I^{\rm up}]=0,
\label{eq:prd_wronskian_rn}
\end{equation}
where $I$ denotes an $\ell\ge2$ axial or polar branch, or the regular radiative dipole $a_d$ of Eq.~\eqref{eq:prd_dipole_vector_master}. Continued-fraction or spectral methods can evaluate these roots numerically \cite{Leaver:1990zz}. Equation~\eqref{eq:prd_wronskian_rn} defines the spectrum implicitly because generic RN quasinormal frequencies have no elementary closed form.

The scalar branch has a distinct long-range asymptotic condition. Expanding Eq.~\eqref{eq:prd_scalar_potential} gives
\begin{equation}
V_{s,\ell}(r)
=
m_G^2-\frac{2\mathcal M m_G^2}{r}
+\frac{\ell(\ell+1)+\mathcal Q^2m_G^2}{r^2}
+\Order(r^{-3}).
\label{eq:prd_scalar_asymptotic_potential}
\end{equation}
The $1/r$ term produces a Coulomb-type logarithmic phase and must remain in the spectral problem. The radial wave number is
\begin{equation}
k_s(\omega)=\sqrt{\omega^2-m_G^2}
\label{eq:prd_scalar_k}
\end{equation}
with its branch obtained by analytic continuation from $k_s>0$ for real $\omega>m_G$, with $k_s/\omega\to1$ for large $|\omega|$ on the outgoing sheet. The radiative outgoing solution has the large-$r$ form
\begin{equation}
\Phi_{\rm out}
=A_{\rm out}e^{+ik_s r_*}
 r^{\,i\mathcal M m_G^2/k_s}
\left[1+\Order(r^{-1})\right].
\label{eq:prd_scalar_outgoing}
\end{equation}
Equivalently, since $r_*=r+2\mathcal M\ln r+\Order(1)$,
\begin{equation}
\Phi_{\rm out}
=A_{\rm out}'e^{+ik_s r}
 r^{\,i\mathcal M(2\omega^2-m_G^2)/k_s}
\left[1+\Order(r^{-1})\right].
\label{eq:prd_scalar_outgoing_r}
\end{equation}
This is the standard long-range massive-field phase structure \cite{Dolan:2007mj,Koyama:2000hj}. A quasi-bound mode is not classified by the shorthand $|\omega|<m_G$ when $\omega$ is complex. The appropriate decay parameter is
\begin{equation}
\kappa_s(\omega)=\sqrt{m_G^2-\omega^2}
\label{eq:prd_scalar_kappa}
\end{equation}
with the branch chosen so that $\operatorname{Re}\kappa_s>0$. The decaying boundary condition is
\begin{equation}
\Phi_{\rm qb}
=A_{\rm qb}e^{-\kappa_s r_*}
 r^{\,\mathcal M m_G^2/\kappa_s}
\left[1+\Order(r^{-1})\right].
\label{eq:prd_scalar_qb}
\end{equation}
The nonzero constants $A_{\rm out}$, $A_{\rm out}'$, and $A_{\rm qb}$ fix the corresponding mode normalizations.
Together with the ingoing horizon condition, Eqs.~\eqref{eq:prd_scalar_outgoing} and \eqref{eq:prd_scalar_qb} define, respectively, the radiative QNM and quasi-bound spectral problems by analytic continuation. The same construction includes the monopole $\ell=0$, whose frequency is not set by the photon-orbit eikonal formula.

The scalar potential determines the stability condition. Outside the outer horizon,
\begin{equation}
\frac{f'}{r}=\frac{2(\mathcal M r-\mathcal Q^2)}{r^4}>0,
\label{eq:prd_fprime_positive}
\end{equation}
where the last inequality follows from $r\ge r_+$ and $\mathcal M^2>\mathcal Q^2$. Hence $V_{s,\ell}(r)\ge0$ for every $\ell\ge0$ when $m_G^2\ge0$. If a mode with $\operatorname{Im}\omega>0$ satisfied the QNM boundary conditions, multiplication of the radial equation by $\Phi^*$ and integration over $r_*$ would give
\begin{equation}
\omega^2\int|\Phi|^2dr_*
=\int\left(\left|\frac{d\Phi}{dr_*}\right|^2+V_{s,\ell}|\Phi|^2\right)dr_*.
\label{eq:prd_scalar_energy}
\end{equation}
The right-hand side is real and nonnegative. Its reality requires an unstable frequency to be purely imaginary, which would make the left-hand side negative. No exponentially growing scalar QNM therefore exists for $m_G^2\ge0$. The gravito-vector subsystem has the mode stability of the subextremal RN perturbation problem \cite{Moncrief:1974gw,Moncrief:1974ng,Moncrief:1975sb}. The exceptional axial dipole is separately stable because $V_{v1}>0$ outside the horizon, while its metric component is a constrained response rather than an independent radiative degree of freedom \cite{Burko:1999jy,Burton:2020wnj}.

Where the Wentzel--Kramers--Brillouin (WKB) approximation applies to a single potential barrier, the leading resonance condition is
\begin{equation}
\omega_n^2=V_0-i\left(n_o+\frac12\right)\sqrt{-2V_0^{(2)}},
\label{eq:prd_wkb}
\end{equation}
where $n_o$ is the overtone number and $V_0^{(2)}$ is the second $r_*$ derivative at the potential maximum \cite{Schutz:1985km,Iyer:1986np,Kokkotas:1988fm}. This formula is not used for the scalar monopole. In the high-$\ell$ limit the dominant massless potentials approach $f\ell(\ell+1)/r^2$, so their maximum approaches the unstable null orbit
\begin{equation}
r_c=\frac{3\mathcal M+\sqrt{9\mathcal M^2-8\mathcal Q^2}}{2}.
\label{eq:prd_rc}
\end{equation}
The orbital frequency and radial Lyapunov exponent are
\begin{align}
\Omega_c&=\sqrt{\frac{r_c-\mathcal M}{2r_c^3}},
\label{eq:prd_omega_lambda}
\\
\Lambda_c&=\sqrt{\frac{(r_c-\mathcal M)(2r_c-3\mathcal M)}{2r_c^4}},
\label{eq:prd_lyapunov}
\end{align}
which give
\begin{equation}
\omega_{\ell n_o}=\ell\Omega_c-i\left(n_o+\frac12\right)\Lambda_c+\mathcal O(\ell^0).
\label{eq:prd_eikonal}
\end{equation}
This geometric limit is shared by RN-like systems and does not by itself distinguish MOG \cite{Cardoso:2008bp}. The MOG charge--mass relation fixes the change in the eikonal scale. At fixed source mass $M$, expansion about $\alpha=0$ gives
\begin{align}
\Omega_c={}&\frac{1}{3\sqrt{3}G_{\mathrm N}M}
\left[1-\frac{5}{6}\alpha+\frac{157}{216}\alpha^2+\mathcal O(\alpha^3)\right],
\label{eq:prd_omega_small_alpha}
\\
\Lambda_c={}&\frac{1}{3\sqrt{3}G_{\mathrm N}M}
\left[1-\frac{17}{18}\alpha+\frac{7}{8}\alpha^2+\mathcal O(\alpha^3)\right].
\label{eq:prd_lambda_small_alpha}
\end{align}
Thus the leading MOG eikonal spectrum is
\begin{multline}
G_{\mathrm N}M\,\omega_{\ell n_o}=\frac{1}{3\sqrt3}\Bigg\{\ell\left[1-\frac56\alpha+\frac{157}{216}\alpha^2\right]
\\
-i\left(n_o+\frac12\right)\left[1-\frac{17}{18}\alpha+\frac78\alpha^2\right]\Bigg\}+\mathcal O(\alpha^3,\ell^0).
\label{eq:prd_eikonal_small_alpha}
\end{multline}
This expansion follows from the MOG background relation and does not introduce a new QNM branch. The scalar has the same leading photon-orbit behavior only when $\ell(\ell+1)/r_c^2\gg m_G^2$.

The vector mass enters the propagation problem after the local strong-field solution is obtained. At a matching radius satisfying $\bar G M\ll r_m\ll\mu^{-1}$, the vector-led near-zone solution separates into asymptotic metric and vector components. In the far zone, a transverse vector wave of real carrier frequency $\omega_R$ has
\begin{align}
k_v^2&=\omega_R^2-\mu^2,
\label{eq:prd_proca_dispersion}
\\
v_{g,v}&=\sqrt{1-\frac{\mu^2}{\omega_R^2}}.
\label{eq:prd_vector_group_velocity}
\end{align}
Its propagation delay relative to a null signal across distance $D$ is
\begin{equation}
\begin{split}
\Delta t_v&=D\left[\frac{1}{\sqrt{1-\mu^2/\omega_R^2}}-1\right]
\\
&=\frac{D\mu^2}{2\omega_R^2}+\mathcal O\left(\frac{D\mu^4}{\omega_R^4}\right).
\end{split}
\label{eq:prd_vector_delay}
\end{equation}
The expansion parameters $\mu\bar G M$ and $\mu/\omega_R$ control different approximations. Over a large propagation distance, the vector wave can accumulate a finite phase correction even when the local mass correction is negligible. This vector-wave delay does not by itself imply a delayed displacement of the optical critical curve, which depends on the metric along the null rays.

\section{Scalar critical-curve invariance and a stationary optical benchmark}
\label{sec:photon_ring}

The effect of the physical metric perturbations on the critical set of null rays follows from the null Hamiltonian
\begin{equation}
H(x,p)=\frac12g^{\mu\nu}p_\mu p_\nu=0.
\label{eq:prd_null_H}
\end{equation}
With $h_{\mu\nu}=\widehat h_{\mu\nu}+\psi\bar g_{\mu\nu}$, the expansion is $H=H_0+\epsilon H_1+\mathcal O(\epsilon^2)$, where
\begin{align}
H_0&=\frac12\bar g^{\mu\nu}p_\mu p_\nu,
\label{eq:prd_H0}
\\
H_1&=-\frac12\widehat h^{\mu\nu}p_\mu p_\nu-\psi H_0.
\label{eq:prd_H1_split}
\end{align}
The second term is the complete scalar contribution. On the background null shell $H_0=0$,
\begin{equation}
H_1^{(s)}=-\psi H_0=0.
\label{eq:prd_scalar_H_zero}
\end{equation}
The scalar-only metric is
\begin{equation}
g_{\mu\nu}^{(s)}=(1+\epsilon\psi)\bar g_{\mu\nu},
\label{eq:prd_scalar_conformal_metric}
\end{equation}
so the Hamiltonian flow on $H_0=0$ differs from the background flow only by a reparametrization. The unparametrized null geodesics are unchanged. The observable compares the same observer worldline and observation event in the two spacetimes, using conformally corresponding orthonormal tetrads,
\begin{equation}
e_{(a)}{}^\mu=(1-\epsilon\psi/2)\bar e_{(a)}{}^\mu.
\label{eq:prd_scalar_observer_identification}
\end{equation}
This comparison assumes a smooth positive conformal factor on the relevant ray domain that preserves the causal boundary used to define capture. The common local tetrad rescaling leaves angular ratios of screen momentum components unchanged. With this observer prescription, the scalar contribution to every linear critical-curve multipole vanishes,
\begin{align}
\mathcal T_{n,s}^{\ell m}(\omega)&=0,
\label{eq:prd_scalar_transfer_zero}
\\
\delta R_n^{(s)}&=0
\label{eq:prd_scalar_screen_zero}
\end{align}
for all $n,\ell,m$. In particular,
\begin{equation}
\delta R_0^{(s,\ell=0)}=0.
\label{eq:prd_no_scalar_breathing}
\end{equation}
The scalar monopole is a dynamical field mode, but it does not cause the geometric critical curve to expand or contract at first order in this reduced MOG sector. This result holds within the linear and near-zone approximations without an adiabatic assumption. It concerns the angular critical curve for the identified observers, not the motion of freely falling material observers selected by identical initial data, a proper-length screen scale, or an accretion brightness profile. A scalar-induced change of an observer's motion would require a separate calculation of aberration and is not included in Eq.~\eqref{eq:prd_scalar_transfer_zero}.

The gravito-vector perturbation can change the null trajectories. In phase-space coordinates $z^A=(x^\mu,p_\mu)$, a background critical ray is denoted by $\bar z^A(\lambda,\varphi)$ and labeled by screen azimuth $\varphi$. Its first-order displacement satisfies
\begin{equation}
\frac{d}{d\lambda}\delta z^A=\mathcal A^A{}_B\delta z^B+s^A,
\label{eq:prd_phase_linear}
\end{equation}
with
\begin{align}
\mathcal A^A{}_B&=\Omega^{AC}\left.\frac{\partial^2H_0}{\partial z^C\partial z^B}\right|_{\bar z},
\label{eq:prd_phase_matrix}
\\
s^A&=-\frac12\Omega^{AB}\left.\frac{\partial(\widehat h^{\mu\nu}p_\mu p_\nu)}{\partial z^B}\right|_{\bar z}.
\label{eq:prd_phase_source}
\end{align}
If $\mathcal U^A{}_B(\lambda,\lambda')$ is the fundamental matrix of the homogeneous variational equation, then
\begin{equation}
\begin{split}
\delta z^A(\lambda)={}&\mathcal U^A{}_B(\lambda,\lambda_o)\delta z^B(\lambda_o)
\\
&+\int_{\lambda_o}^{\lambda}d\lambda'\,\mathcal U^A{}_B(\lambda,\lambda')s^B(\lambda').
\end{split}
\label{eq:prd_phase_solution}
\end{equation}
This expression retains the full time dependence without assuming conservation of $p_t$ in a dynamical metric.

For observer tetrad $e_{(a)}{}^\mu$, the screen coordinates are chosen as
\begin{align}
X&=-r_o\frac{p_{(\varphi)}}{p_{(r)}},
\label{eq:prd_screen_xy}
\\
Y&=r_o\frac{p_{(\theta)}}{p_{(r)}},
\label{eq:prd_screen_y}
\end{align}
with $p_{(a)}=e_{(a)}{}^\mu p_\mu$. The screen response includes both the tetrad variation and the propagated momentum variation. A symmetric tetrad choice gives $\delta e_{(a)}{}^\mu=-\widehat h^\mu{}_\nu\bar e_{(a)}{}^\nu/2$, up to a local Lorentz rotation that fixes the screen orientation. The critical curve is the intersection of the observer screen with the stable manifold of the unstable photon orbit. If a scalar function $\mathcal C[z_o(R,\varphi)]$ defines that manifold, its first variation gives
\begin{equation}
\delta R(\varphi,t_o)=-\frac{\delta_{\widehat h}\mathcal C}{\partial_R\mathcal C}\bigg|_{R=R_0}.
\label{eq:prd_deltaR_general}
\end{equation}
For completeness, a formal boundary-value representation of the numerator uses an adjoint variational covector $\chi_A$:
\begin{equation}
\begin{split}
\delta_{\widehat h}\mathcal C={}&\delta\mathcal C_{\rm obs}+\delta\mathcal C_{\rm trap}
\\
&+\int_{\lambda_o}^{\lambda_b}d\lambda\,\chi_A\Omega^{AB}
\frac{\partial H_1^{(gv)}}{\partial z^B},
\end{split}
\label{eq:prd_transfer_kernel}
\end{equation}
where $H_1^{(gv)}=-\widehat h^{\mu\nu}p_\mu p_\nu/2$. Here $\lambda_b$ denotes a terminal surface near the trapped set, with the affine direction chosen from the observer toward that surface. If $n_A$ is the normal to its background capture separatrix, the adjoint satisfies
\begin{equation}
\frac{d\chi_A}{d\lambda}=-\chi_B\mathcal A^B{}_A,
\qquad \chi_A(\lambda_b)=n_A.
\label{eq:prd_adjoint_boundary}
\end{equation}
The observer term is $\delta\mathcal C_{\rm obs}=\chi_A(\lambda_o)\delta z_{\rm frame}^A$, and $\delta\mathcal C_{\rm trap}$ is the explicit perturbation of the terminal separatrix condition. The perturbed trapped set determines this terminal condition, and the calculation must take the critical-ray limit consistently. The trapping term cannot be set to zero by assumption. Under a coordinate gauge transformation, $H_1^{(gv)}$ changes by a total derivative along the background geodesic. Transforming both endpoints and the observer tetrad cancels that boundary term. This cancellation checks the pure-gauge response but does not evaluate a transfer coefficient. Equations~\eqref{eq:prd_phase_solution}--\eqref{eq:prd_adjoint_boundary} specify the formal ingredients of a global response calculation of the type required by time-dependent shadow studies \cite{Wang:2019skw,Koga:2022dsu,Pantig:2025eqe}. The present analysis does not evaluate the finite-frequency gravito-vector kernel.

The critical-curve perturbation has the Fourier expansion
\begin{equation}
\delta R(\varphi,t)=\sum_{n=-\infty}^{\infty}\delta R_n(t)e^{in\varphi},
\label{eq:prd_screen_modes}
\end{equation}
with reality condition $\delta R_{-n}=\delta R_n^*$. The relation between mode amplitudes and screen displacements is
\begin{equation}
\delta R_n(\omega)=\sum_{I=g,v}\sum_{\ell m}
\mathcal T_{n,I}^{\ell m}(\omega,\alpha,r_o,\iota)\mathcal A_{I\ell m}(\omega).
\label{eq:prd_transfer_gv}
\end{equation}
There is no scalar term because of Eq.~\eqref{eq:prd_scalar_transfer_zero}. The $n=0$ component changes the mean radius, the $|n|=1$ components translate the center, and the $|n|\ge2$ components change the shape.

\subsection{Local low-frequency limit of the radiative dipole}

The source-free scattering problem takes the axis-aligned $m=0$ vector dipole with $\mathfrak j_d=0$. For real $\omega>0$, the radiative solution of Eq.~\eqref{eq:prd_dipole_vector_master} satisfies
\begin{align}
a_d&=\mathcal T_v e^{-i\omega r_*}\left[1+o(1)\right],
&&r_*\to-\infty,
\label{eq:prd_dipole_scatter_horizon}
\\
a_d&=e^{-i\omega r_*}+\mathcal R_v e^{+i\omega r_*}+o(1),
&&r_*\to+\infty.
\label{eq:prd_dipole_scatter_infinity}
\end{align}
The unit incident coefficient fixes the normalization of the canonical vector amplitude. Any physical incident amplitude multiplies all reconstructed fields. Equation~\eqref{eq:prd_dipole_metric_integral} gives the metric with the asymptotic angular frame fixed as specified there. A superposition of these real-frequency solutions can form a causal packet. An isolated QNM has a radial profile that grows at infinity and cannot be inserted directly into the convergent real-frequency integral.

The vector equation also determines the low-frequency limit. Its horizon-regular static solution, normalized to have leading behavior $r^2$ at large radius, is
\begin{equation}
A_0(r)=r^2-3\mathcal Q^2+\frac{2\mathcal Q^4}{\mathcal M r}.
\label{eq:prd_dipole_static_profile}
\end{equation}
Direct substitution gives
\begin{equation}
(fA_0')'-\left(\frac{2}{r^2}+\frac{4\mathcal Q^2}{r^4}\right)A_0=0.
\label{eq:prd_dipole_static_identity}
\end{equation}
For a fixed subextremal background, the second independent solution is logarithmically singular at the horizon. A solution regular at the horizon and decaying at infinity would obey
\begin{equation}
\int_{r_+}^{\infty}\left[f|A'|^2+
\left(\frac{2}{r^2}+\frac{4\mathcal Q^2}{r^4}\right)|A|^2\right]dr=0,
\label{eq:prd_dipole_static_nohair}
\end{equation}
after integration by parts, so it must vanish. There is no nonzero, asymptotically decaying static vector dipole in this zero-angular-momentum sector.

For $\omega\mathcal M\ll1$, the horizon-regular and outer solutions match in the overlap region $\mathcal M\ll r\ll\omega^{-1}$. At leading order the outer equation is the flat-space $\ell=1$ radial equation. With $z=\omega r$, its solution with equal unit incoming and outgoing leading coefficients is
\begin{equation}
-2zj_1(z)=-\frac{2}{3}z^2+\Order(z^4),
\label{eq:prd_dipole_outer_match}
\end{equation}
where $j_1$ is the spherical Bessel function. The coefficient follows from the exact identity $zj_1(z)=-\cos z+\sin z/z$. Matching the growing $r^2$ term fixes the inner amplitude. The absence of a static decaying solution excludes a threshold resonance in the vector problem. With the tortoise phase convention in Eqs.~\eqref{eq:prd_dipole_scatter_horizon} and \eqref{eq:prd_dipole_scatter_infinity},
\begin{equation}
\lim_{\omega\to0^+}\frac{a_d(r,\omega)}{\omega^2}
=-\frac{2}{3}A_0(r)
\label{eq:prd_dipole_lowfreq_vector}
\end{equation}
at fixed $r>r_+$. Subleading terms can contain long-range logarithmic phases. This result does not assume a Taylor expansion of the full scattering solution at infinity.

The metric integral has an equally regular local limit. In the near and overlap regions its integrand is controlled by $\omega^2 A_0(s)/s^4$. The contribution from $s\gtrsim\omega^{-1}$ is of order $\omega^3$, since the scattering amplitudes there are bounded. Splitting the integral at an overlap radius and matching the two regions therefore yields
\begin{equation}
\begin{split}
\lim_{\omega\to0^+}\frac{\widetilde h_v(r,\omega)}{\omega^2}
&=-\frac{8\mathcal Q r^2}{3}\int_r^\infty\frac{A_0(s)}{s^4}\,ds
\\
&=-\frac{8\mathcal Q}{3}\left(r-\frac{\mathcal Q^2}{r}
+\frac{\mathcal Q^4}{2\mathcal M r^2}\right).
\end{split}
\label{eq:prd_dipole_lowfreq_metric}
\end{equation}
For this normalization, the local metric response tends to zero quadratically with frequency. The limit is taken at fixed subextremal charge and fixed radius, not uniformly to spatial infinity or extremality. It is a statement about the reconstructed local metric, not about a finite-frequency optical centroid.

This first-order reconstruction also avoids an auxiliary zero-mode ambiguity. The metric-sector potential
\begin{equation}
V_{g1}=f\left(\frac{2}{r^2}-\frac{6\mathcal M}{r^3}
+\frac{4\mathcal Q^2}{r^4}\right)
\label{eq:prd_dipole_metric_potential}
\end{equation}
satisfies
\begin{equation}
\left(\frac{d^2}{dr_*^2}-V_{g1}\right)\frac1r=0.
\label{eq:prd_dipole_auxiliary_zero_mode}
\end{equation}
Consequently, eliminating an explicit $1/\omega$ from a reconstruction through that second-order operator would not by itself establish a regular inverse at zero frequency. Equations~\eqref{eq:prd_dipole_metric_integral} and \eqref{eq:prd_dipole_lowfreq_metric} use neither that inverse nor an assumption that its zero mode is absent.

In ingoing coordinates, the metric components are
\begin{align}
h_{v\varphi}^{(d)}&=-N_1\widetilde h_v(r,\omega)
\sin^2\theta\,e^{-i\omega v},
\\
h_{r\varphi}^{(d)}&=0,
\qquad h_{\theta\varphi}^{(d)}=0.
\label{eq:prd_dipole_regular_metric_components}
\end{align}
The nonzero angular component is regular at the future horizon and vanishes quadratically at both axes.

\subsection{Stationary optical benchmark}

A separate stationary calculation tests the metric-to-optics normalization without replacing a dynamical capture problem by an instantaneous photon orbit. For an asymptotically nonrotating, stationary axial perturbation, the calculation uses an exterior gauge with $h_{r\varphi}=0$ and a static observer in the asymptotic equatorial plane. The equatorial metric is
\begin{equation}
\begin{split}
ds^2={}&-fdt^2+\frac{dr^2}{f}+r^2d\varphi^2
+2\epsilon h_{t\varphi}^{(d)}(r)dt\,d\varphi .
\end{split}
\label{eq:prd_axial_dipole_metric}
\end{equation}
The exact stationary Killing integrals are $E=-p_t$ and $L=p_\varphi$. With $b=L/E$, the circular-null conditions are $F(r,b)=0$ and $\partial_rF=0$, where
\begin{equation}
F(r,b)=g^{tt}-2bg^{t\varphi}+b^2g^{\varphi\varphi}.
\label{eq:prd_stationary_F}
\end{equation}
At the background critical orbit $b_\pm^{(0)}=\pm b_c$, the first-order radial shift drops out because $\partial_rF=0$. Hence
\begin{equation}
\delta b_+=\delta b_-=\frac{h_{t\varphi}^{(d)}(r_c,\pi/2)}{f(r_c)}.
\label{eq:prd_dipole_shift}
\end{equation}
The half-width between the equatorial endpoints is unchanged while their midpoint translates. The infinitesimal Kerr--Newman family provides an exact linearized electrovac solution with a stationary angular-momentum perturbation on which to test this formula \cite{Newman:1965my}. Its spin length is $a_{\rm rot}$, with $|a_{\rm rot}|\ll\mathcal M$. In the harmonic and charge conventions used here,
\begin{align}
h_t&=\frac{a_{\rm rot}}{N_1}
\left(\frac{2\mathcal M}{r}-\frac{\mathcal Q^2}{r^2}\right),
\\
a_d&=-\frac{\mathcal Q a_{\rm rot}}{N_1r},
\qquad h_r=0.
\label{eq:prd_KN_benchmark_fields}
\end{align}
Substitution into Eq.~\eqref{eq:prd_dipole_angular_momentum} gives
\begin{equation}
\mathfrak j_d=-\frac{6\mathcal M a_{\rm rot}}{N_1},
\label{eq:prd_KN_benchmark_charge}
\end{equation}
which fixes the relation to the geometric angular momentum $\mathcal J=\mathcal M a_{\rm rot}$. These fields also satisfy the angular Maxwell equation and the static vector equation with source $-f\mathcal Q\mathfrak j_d/r^4$. Since $h_{t\varphi}=-a_{\rm rot}(2\mathcal M/r-\mathcal Q^2/r^2)\sin^2\theta$, the equatorial impact-parameter midpoint shift is
\begin{equation}
\delta C_b^{\rm KN}
=-a_{\rm rot}\frac{2\mathcal M/r_c-\mathcal Q^2/r_c^2}{f(r_c)}.
\label{eq:prd_stationary_centroid_benchmark}
\end{equation}
For the asymptotic screen convention $X=-b$, the screen midpoint shifts by $-\delta C_b^{\rm KN}$. In the neutral limit $r_c=3\mathcal M$ and $f(r_c)=1/3$, giving $\delta C_b^{\rm KN}=-2a_{\rm rot}$ and the opposite screen shift. This benchmark checks the signs, vector normalization, angular-momentum integration constant, and stationary optical map. A stationary radial angular gauge transformation leaves $h_{t\varphi}$ unchanged. A rotating asymptotic frame would alter the observer prescription and is excluded. The Kerr--Newman geometry is regular at the future horizon in ingoing rotating coordinates, even though the displayed exterior gauge is not a horizon chart.

The benchmark has $\mathfrak j_d\ne0$ and therefore differs from the zero-angular-momentum scattering experiment. Its neutral limit need not vanish because it retains spin. It does not establish a nonzero radiative centroid transfer. For a time-dependent scattering mode or a causal ringdown signal, $E=-p_t$ is not conserved, and Eq.~\eqref{eq:prd_dipole_shift} cannot be applied instantaneously. The local inequality $|\omega|/\Lambda_c\ll1$ alone does not control propagation to a distant observer or the critical-ray boundary condition. A vector-to-image prediction therefore requires evaluation of the global boundary-value problem, including its observer and trapping terms. No finite-frequency scattering-to-centroid coefficient is claimed here.

A finite vector mass changes far-zone propagation. Beyond the matching region, the vector component acquires the phase $e^{ik_vD}$ with $k_v=\sqrt{\omega^2-\mu^2}$, while freely propagating metric radiation remains null. A delay in the vector signal is therefore not the same as a delayed optical centroid. For vector dispersion to delay the image displacement, the far-zone vector field would need to induce a metric perturbation along the optical path or at the source. This effect lies beyond the leading local calculation. The established optical results are the vanishing scalar response for the specified observers and the separate stationary benchmark, not a delayed radiative image displacement.

\section{Consistency checks and limiting cases}
\label{sec:consistency}

At $\alpha=0$, the background parameters satisfy $\mathcal Q=0$, $\bar G=G_{\mathrm N}$, $r_+=2G_{\mathrm N}M$, and $r_c=3G_{\mathrm N}M$. The two gravito-vector master families separate into electromagnetic and gravitational Schwarzschild sectors. At fixed $G_{\mathrm N}$, $M$, and exterior radius $r$, their potentials have the exact limits
\begin{equation}
\lim_{\alpha\to0}V_1^{(\pm)}=f_{\rm Schw}\frac{\ell(\ell+1)}{r^2},
\label{eq:prd_em_limits}
\end{equation}
\begin{equation}
\lim_{\alpha\to0}V_2^{(-)}=f_{\rm Schw}\left[\frac{\ell(\ell+1)}{r^2}-\frac{6G_{\mathrm N}M}{r^3}\right],
\label{eq:prd_rw_limit}
\end{equation}
and
\begin{equation}
\begin{split}
\lim_{\alpha\to0}V_2^{(+)}={}&\frac{2f_{\rm Schw}}{r^3(nr+3G_{\mathrm N}M)^2}
\\
&\times\Bigl[n^2(n+1)r^3+3n^2G_{\mathrm N}Mr^2
\\
&\qquad+9n(G_{\mathrm N}M)^2r+9(G_{\mathrm N}M)^3\Bigr],
\end{split}
\label{eq:prd_zerilli_limit}
\end{equation}
Here $f_{\rm Schw}=1-2G_{\mathrm N}M/r$, and Eq.~\eqref{eq:prd_zerilli_limit} is the Zerilli potential. These limits check the signs in Eqs.~\eqref{eq:prd_W}--\eqref{eq:prd_polar_potential_2}.

The Schwarzschild background limit differs from the pure-GR perturbative limit. Equation~\eqref{eq:prd_scalar_KG} remains when $\alpha=0$ because $G$ is still a dynamical scalar in the reduced action. Pure vacuum GR is recovered only as $\alpha\to0$ on the consistent subspace $a_\mu=0$ and $\psi=0$. The background parameter limit alone does not remove the additional degrees of freedom.

The field equations also provide a direct check of the field redefinition. In four dimensions, a conformal perturbation satisfies Eq.~\eqref{eq:prd_conformal_Einstein}, and the covariant Maxwell stress tensor varies according to Eq.~\eqref{eq:prd_conformal_Maxwell}. The conformal terms in Eq.~\eqref{eq:prd_decoupled_vector} cancel with coefficients $-1-1+2=0$. The diagonalization therefore holds for any gauge choice and multipole. The scalar critical-curve response vanishes because the scalar Hamiltonian perturbation is proportional to the null constraint.

The scalar kinetic coefficient is positive in the domain considered here. The scalar-tensor part of the action has the form $F(G)R/(16\pi)-Z(G)(\nabla G)^2/2$, with $F=G^{-1}$ and $Z=G^{-3}$. After conformal diagonalization, the kinetic coefficient is proportional to
\begin{equation}
\frac{1}{G^2}\left(1+\frac{3}{16\pi}\right)>0
\label{eq:prd_scalar_kinetic}
\end{equation}
for $G>0$. A nonnegative effective mass squared $m_G^2$, defined in Eq.~\eqref{eq:prd_scalar_mass}, then excludes the tachyonic instability identified by the scalar potential.

The near-zone and far-zone vector approximations use distinct small parameters. Neglecting the local Proca term requires $\mu\bar G M\ll1$, while expanding the far-zone group delay requires $\mu/\omega_R\ll1$. The accumulated phase correction $D\mu^2/\omega_R$ can remain finite. The massless local theory has two transverse vector polarizations. A finite-$\mu$ strong-field treatment would also contain a longitudinal Proca mode and require a background corrected at the same order. The present analysis does not address that mode.

In mass units, dimensional analysis gives $[G]=-2$, $[M]=[\mu]=[\omega]=1$, $[r]=-1$, $[\phi_\mu]=1$, and $[B_{\mu\nu}]=2$. The potential enters the Lagrangian as $V_G/G^3$, so $[V_G]=-2$ and $[V_G''(\bar G)]=2$, as required for $m_G^2$. The effective lengths $\mathcal M$ and $\mathcal Q$ both have mass dimension $-1$. These dimensions are consistent with the lapse, master potentials, spectral conditions, and propagation relations.

\section{Conclusion}
\label{sec:conclusion}

The reduced strong-field MOG theory separates scalar and gravito-vector perturbations through $\widehat h_{\mu\nu}=h_{\mu\nu}-\psi\bar g_{\mu\nu}$. The scalar remains dynamical and is stable under the stated boundary conditions when $m_G^2\ge0$, but its metric perturbation is conformal. For fixed observer worldlines and conformally corresponding tetrads, every linear scalar-to-critical-curve response therefore vanishes, including monopole breathing. This result assumes a smooth positive conformal factor that preserves the capture boundary. It establishes that a scalar polarization affecting material test particles need not change the angular critical curve, so the existence of a field mode alone does not determine its optical signature.

The charged axial dipole instead induces a constrained metric perturbation. Separating the angular-momentum constant from the radiative vector amplitude gives a first-order reconstruction that is regular at the future horizon and on the sphere, without division by frequency. For a unit-incident canonical vector amplitude with zero angular momentum, the local metric response vanishes as $\omega^2$ at fixed radius and fixed subextremal background. The explicit charge-dependent coefficient supplies a low-frequency check for perturbation calculations and rules out a nonzero static limit of this normalized scattering response. It does not determine a finite-frequency image displacement.

The infinitesimal Kerr--Newman benchmark independently checks the stationary optical normalization and recovers the neutral impact-parameter midpoint shift $\delta C_b=-2a_{\rm rot}$. Its nonzero angular momentum distinguishes it from the scattering problem. A finite vector mass can also delay vector propagation without establishing an optical delay. Neither stationary rotation nor vector dispersion can replace the global null-propagation calculation required for a radiative centroid signal.

These results connect the field perturbations to a specified optical observable and distinguish scalar angular invariance from vector-induced metric changes. Within the reduced action and observer prescription, the calculation excludes linear scalar breathing and supplies a regular vector-induced metric perturbation for further lensing analysis. This provides a field-equation-based test of the proposed breathing and delayed-vector sequence \cite{Lobos:2026jyp}. The results concern the geometric critical curve, not an accretion brightness profile or the full STVG theory. Predicting a vector-induced image signal still requires a specified excitation and a global calculation that includes the observer frame and critical-ray boundary conditions.

\begin{acknowledgments}
N.J.L. Lobos and E.T. Rodulfo gratefully acknowledge De La Salle University and the DLSU Theoretical Physics Group for their institutional support. Furthermore, we extend our sincere gratitude to the Department of Science and Technology – Accelerated Science and Technology Human Resource Development Program (DOST-ASTHRDP) for their generous and continuous support of our research endeavors.
\end{acknowledgments}

\section*{Data Availability}
No datasets reported in this article.

\bibliographystyle{apsrev4-2}
\bibliography{ref}

\end{document}